\documentclass[11pt,twoside]{article}

\usepackage{asp2006}
\usepackage{epsf}
\usepackage{psfig}
\usepackage{lscape}
\usepackage{graphicx}

\begin{document}
\title{Sensitivity of p modes for constraining velocities of 
microscopic diffusion of the elements}
\author{Orlagh L. Creevey}
\affil{Instituto de Astrof\'isica de Canarias, 
C/ V\'ia Lactea s/n, La Laguna 38205, Tenerife, Spain: Email: orlagh@iac.es}
\author{Savita Mathur}   
\affil{Indian Institute of Astrophysics, Koramangala, Bangalore 560034, India}   
\author{Rafael A. Garc\'ia}
\affil{Laboratoire AIM, CEA/DSM-CNRS-Universit\'e Paris Diderot; 
CEA, IRFU, SAp, F-91191, Gif-sur-Yvette, France}

\begin{abstract} 
Conventional astrophysical observations have failed to provide stringent constraints on physical processes operating in the interior of the stars. However, satellite missions now promise a solution to these problems by providing long-term high-quality continuous data which will allow the application of seismic techniques. With this in mind, and using the Sun as our astrophysical laboratory, our aim is to determine if Corot- and Kepler-like asteroseismic data can constrain physical processes like microscopic diffusion. We test to what extent can the observed atmospheric abundances coupled with p-mode frequencies safely distinguish between stellar initial chemical composition and diffusion of these elements. We present some preliminary results of our analysis.
\end{abstract}



\section{Introduction \& Method}

If we have the following observations for a solar-type star:
a set of p-mode frequencies, measured abundances, and an identified g mode,
we can investigate how these observations may constrain some of the 
physical processes occuring in the interior of a star.
Suppose that the stellar model is a main sequence 1 M$_{\odot}$ star, and that
we can calibrate the age of the star using the radius and luminosity.
Then the {\it free} parameters that remain to be fitted in the star are
the initial mass fractions of hydrogen and metals $X_0$ and $Z_0$, 
and the velocities of diffusion of these elements $V_X$, $V_Y$ and $V_Z$, 
(assuming that other microscopic processes may not be ``observable''), 
where $Y$ denotes helium.
Let's denote these set of parameters as {\bf P}.
Using {\bf P} we can calculate a stellar model which will give us
the {\it expected observables}, {\bf B} (expected p-mode frequencies, 
abundances etc.).
If we have a set of observations {\bf O}, then 
in order to find the {\bf P} that most adequately reproduce
{\bf O}, we can use the following equation iteratively, until we 
make {\bf O} as close as possible to {\bf B}:
\begin{equation}
\boldmath \delta {\bf P = V}{\bf {\tilde W}^{-1}} {\bf U} \frac{{\bf O - B}}
{\epsilon}.\unboldmath
\label{eqn:dp}
\end{equation}
\boldmath$\delta${\bf P }\unboldmath are the parameter corrections to make
to the initial guess {\bf P$_0$}, \boldmath$\epsilon$ \unboldmath are the 
measurement errors, and {\bf UWV$^T$} is the singular value decomposition of 
the sensitivity matrix, which can be written as  
\boldmath$\frac{\partial B}{\partial P}/\epsilon $ \unboldmath and 
$\tilde{{\bf W}^{-1}}$ has some zero values to stabilize the inversion
(see \cite{cre08,cre09}).

Once {\bf P} have been found, 
then the 
theoretical uncertainties \boldmath$\sigma$({\bf P}) \unboldmath 
can be calculated for each $j = 1, 2, ... N$ parameter
\begin{equation}
\boldmath
\sigma_j^2({\bf P}) = \sum_k^N \frac{{\bf V}_{jk}^2}{{\bf \tilde {W}}_k^2}.
\unboldmath
\label{eqn:one}
\end{equation}
The partial derivatives were calculated using the CESAM code 
{\it Code d'Evolution Stellaire Adaptatif et Modulaire} \citep{mor97} 
and ADIPLS \citep{jcd07}. 
We took the standard  solar model as a reference using the 
abundances of \citet{gn93}, and calibrated these models 
in radius and luminosity 
to an accuracy of $\sim10^{-2}$ as described in \citet{mat07}.
The reference parameters for the models are $(X_0, Z_0) = (0.708, 0.0195)$
and $(V_X, V_Y, V_Z) = (2.307, -7.063, -5.635) \times 10^{-9}$ cm s$^{-1}$. 


\begin{figure}
\center{\includegraphics[width=0.8\textwidth]
{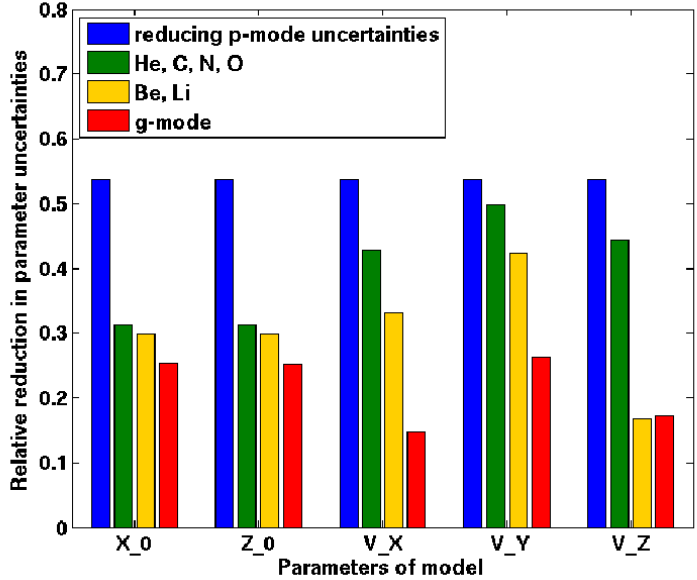}}
\caption{The relative uncertainties as a fraction of the initial values of 
3\%, 24\%, 189\%, 217\%, 422\% (details in text)  as
extra observables are included.\label{fig:poster1}}
\end{figure}

\section{Results and Discussion}

\subsection{Impact of each observable on the uncertainties}
Using only p-mode frequencies to determine the parameter uncertainties 
(using Eq.~\ref{eqn:one}) the diffusion coefficients $V_X$, $V_Y$ and $V_Z$ 
can not be disentangled.  
Their uncertainties remain too large
(189\%, 217\% and 422\% respectively) when we assume that we 
have {\it only} these 
observations with errors of the order of 1.3$\mu$Hz,
meanwhile the values for $\sigma(X_0)$ and $\sigma(Z_0)$ are 3\% and 24\%.  
However, to show the effect that each of the observations has on the 
determination of the parameters, 
Figure 1 shows how the uncertainties in each of the parameters reduce 
relative to these values, 
as extra observables such as abundances and       
g modes are included.  
Each of the bars represents from left to right the following 
additions to the set of observations:
\begin{itemize}
 \item {\sl Blue} reduce the errors on the p-mode frequencies by                          almost a factor of 2 (from 1.3 $\mu$Hz –-- 0.7$\mu$Hz)
 \item {\sl Green} include abundance measurements of He, C, N and O
 \item {\sl Yellow} include abundance measurements of Li and Be 
            ($\epsilon$ = 0.1 dex for all abundances)
  \item{\sl Red} include 1 identified g mode
	\end{itemize}
After including all of these observations, the final parameter 
uncertainties are: 1\%, 6\%, 28\%, 57\% and 72\% respectively.

It is interesting to note the various effects of adding in new observables.
Firstly, reducing all of the uncertainties by almost a factor of 2 should
result in a corresponding reduction in the uncertainties by this same amount. 
The blue bars clearly indicate that this is the case.
When we include the abundance measurements of 
He, C, N, O with an observational error of 0.1 dex, the uncertainties
in both $X_0$ and $Z_0$ reduce by almost a factor of two (green).
These same abundances have little effect for constraining the diffusion 
velocities.
However, including Be and Li measurements has a significant impact on the 
determination of the metal diffusion velocity $V_Z$ (yellow), 
reducing its uncertainty
to 72\% ($\sim$20\% of its original value).
Finally, the inclusion of one identified g mode has most impact on
constraining $V_X$ and then $V_Y$ (red).

\begin{figure}
\center{\includegraphics[width=0.8\textwidth,height=0.35\textheight]{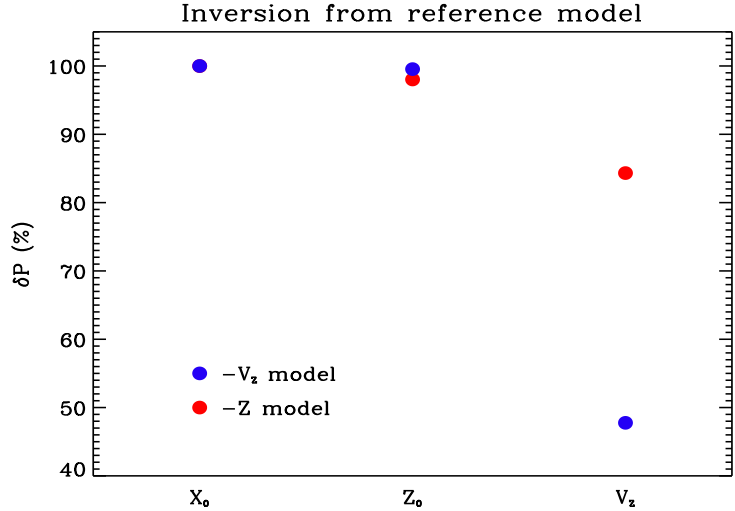}}
\caption{The parameter changes \boldmath $\delta${\bf P} \unboldmath
needed to make to {\bf P$_0$} to correctly fit the new observations.
These new observations were generated using a model with a lower $Z_0$ (red)
and a lower $V_Z$ (blue).
\label{fig:two}}
\end{figure}

\subsection{Recovery of original parameters}
The final uncertainties may be sufficiently low that the observations
do contain enough  information to be able to distinguish safely between
initial metal mass fraction $Z_0$ and diffusion of this $V_Z$
(our original scientific question).
Assume that we have two new sets of observations, and that these {\bf O} 
really come from a solar model with 1) lower $Z_0$ and 2) lower $V_Z$.
If we use Eq.~\ref{eqn:dp} to fit the new {\bf O} (separately) while
using the reference set of parameters as {\bf P$_0$} (the initial
guess), 
we will obtain non-zero negative values
of $\delta Z_0$ and $\delta V_Z$ respectively, 
{\it only if} the observations contain sufficient information.
If there is not enough information contained in the observations, 
then  $\delta Z_0\sim\delta V_Z\sim 0$. 

Figure \ref{fig:two} shows the values of \boldmath$\delta$P \unboldmath 
(relative to the original value of {\bf P}, for scaling purposes) 
for some of the parameters of the models, when attempting to fit a set
of observations generated from a model with a reduced $Z_0$ (red) and 
a reduced $V_Z$ (blue).
As the figure shows, some non-zero \boldmath$\delta$P \unboldmath are needed
to account for the new observations.



\begin{itemize}
\item When inverting using the lower $Z_0$ model observations (red), 
$\delta Z_0$ is negative.
This means that we need to reduce the original $Z_0$ value to a lower
one --- consistent with the input model. 
However, we also see a negative contribution to $V_Z$; 
ideally $V_Z$ would not change, because it was not changed in this model, 
just as $X_0$ does not change.  
\item When inverting using the lower $V_Z$ model observations (blue), 
we obtain a much more negative 
$\delta V_Z$ (but no change in $Z_0$), 
indicating a necessary downward revision of only $V_Z$ ---  
consistent with the input model.  
\end{itemize}

These results are preliminary, but encouraging: 
there is some indication in the set of observables that we may be 
able to distinguish between initial metal abundance and 
diffusion of this using abundance measurements and seismic data.

\acknowledgements 
OLC wishes to acknowledge David Salabert for help with MATLAB. 
RAG is very grateful to the IAC for its visitor support.
SM wishes to thank S\'ebastien Couvidat for his help with Fortran.
This research was in part supported by the European Helio- and 
Asteroseismology Network (HELAS), a major international collaboration 
funded by the European Commission's Sixth Framework Programme
and by the CNES/GOLF grant at SAp, CEA, Saclay. 


\end{document}